\documentclass[10pt,letterpaper,copyedit]{article}
\usepackage{osajnl2}
\usepackage{epsf}
\usepackage{graphicx,epsf,subfigure}
\usepackage{amsmath,amsfonts,mathrsfs}

\begin{document}
\title{Digital in-line holography in thick optical systems: application to visualization in pipes}

\author{N. Verrier, S. Co\"etmellec, M. Brunel and D. Lebrun}
\address{ Groupe d'Optique et d'Opto\'electronique, UMR-6614
CORIA, Av. de l'Universit\'e, \\76801 Saint-Etienne du Rouvray cedex,
France }

\email{coetmellec@coria.fr, verrier@coria.fr}

\begin{abstract}
\textbf{In this paper we apply digital in-line holography to image
opaque objects through a thick plano-concave pipe. Opaque fibers and opaque
particles are considered}. Analytical expression of the intensity
distribution in the CCD sensor plane is derived using generalized
Fresnel transform. \textbf{The proposed model has the ability to
deal with various pipe shape and thickness and compensates for the
lack of versatility of classical DIH models. Holograms obtained with
a 12 mm thick plano-concave pipe are then reconstructed using
fractional Fourier transform (FRFT).} This method allows us to get
rid of astigmatism. Numerical and experimental results are
presented.
\end{abstract}


\section{Introduction}

Digital in-line holography (DIH) is a recognized optical technique
for flow measurements in transparent liquid media. This method is
widely used in various domains such as fluid
mechanics~\cite{Pan,Malek} where the flow is seeded with small
particles, or in the biological microscopic imaging
field~\cite{Malkiel,GarciaS}. \textbf{Nevertheless, a transparent
pipe can be viewed as an optical system that can introduce
aberrations such as astigmatism~\cite{Vikram}. These aberrations
make difficult to image objects in the pipe. Compensation for these
unwanted effects has been widely investigated. De Nicola \emph{et
al.}~\cite{DeNicola2001} proposed a numerical method to compensate
for anamorphism. By modifying the chirp function and the spatial
frequency term of the Fresnel integral the authors managed to
compensate for a severe anamorphism brought by a reflexion
diffraction grating. Numerical methods for astigmatism compensation
have been proposed in
Refs.~\cite{DeNicola2001,Grilli2001,DeNicola2002}. Here the authors
used a modified chirp function where two different propagation
distances are considered. Astigmatism can also be compensated for by
optical means. For instance, the authors of~\cite{Crane} used index
matching to minimize astigmatism: the thin-lens like effect of an
ampule filled with contaminants was investigated to determine index
matching parameters for astigmatism compensation. However, the pipe
used to carry a flow must be considered as a thick and cylindrical
optical system.} Recently, an analytical solution of the scalar
diffraction produced by an opaque disk illuminated by an elliptical,
astigmatic and Gaussian beam under Fresnel approximation has been
proposed~\cite{Nicolas}. In this publication, the astigmatism was
introduced and controlled by using a thin plano-convex cylindrical
lens. Using fractional Fourier transformation (FRFT), authors
managed to retrieve a correct image of an opaque disk from
astigmatic hologram. It was shown that FRFT is therefore well suited
for these studies where classical methods such as Fresnel
transform~\cite{Onural} or Wavelet transform~\cite{Onural_2} fail.

The astigmatism introduced by thin lenses is thus well controlled.
However, these DIH models can not be applied to hugger pipes. As a
matter of fact, the optical thickness of these has to be taken into
account.

In this paper, the aim is to apply DIH to thick pipe systems. In the
first part of the paper, the intensity distribution, in the CCD
plane, of the diffracted field produced by an opaque object located
in a pipe is calculated. \textbf{A method of Gaussian functions
superposition is used to describe the object function.} In the
second part, definition of the FRFT is recalled and we demonstrate
its ability to reconstruct holograms recorded in such systems.
Finally simulations and experiments are performed to illustrate our
results.

\section{In-line Holography through pipes}

\begin{figure}[t]
\centering
\subfigure[]{\includegraphics*[width=15cm]{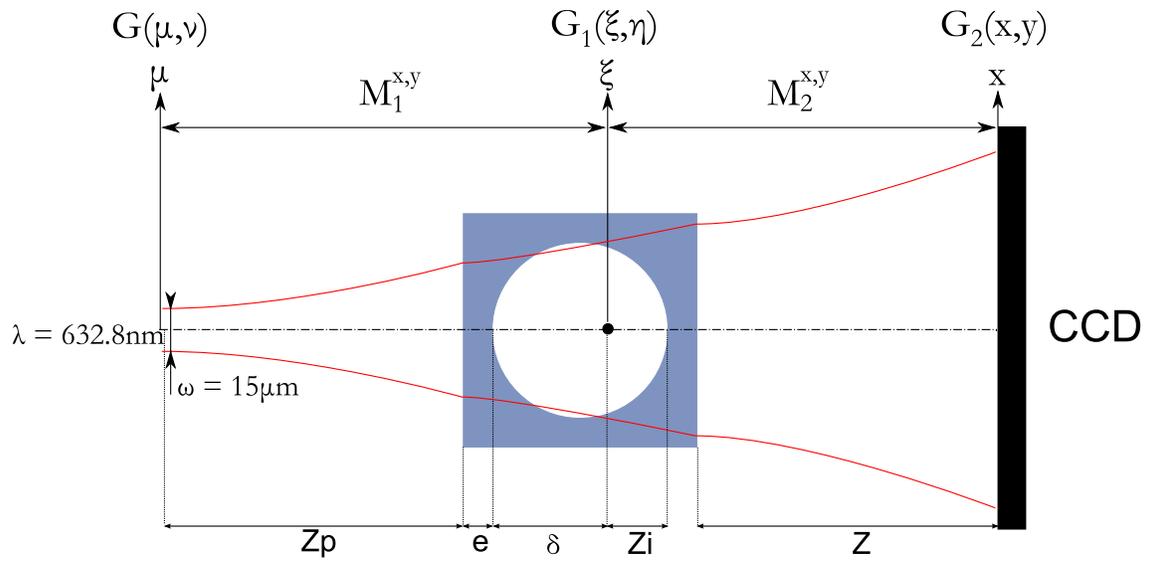}}\quad
\subfigure[]{\includegraphics*[width=6cm]{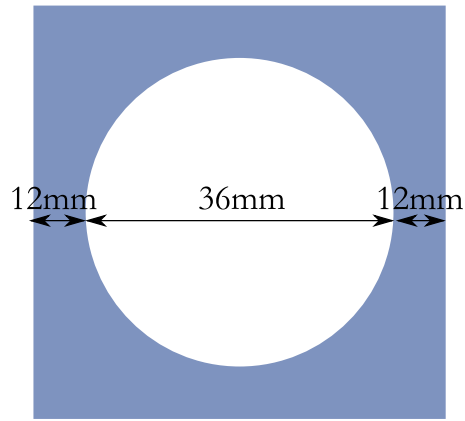}}
\caption{a) Schematic representation of the optical set-up (not to
scale). Definition of the numerical and experimental
parameters. b) Close-up of the pipe used in simulations and experiments}\label{fig1}
\end{figure}

Holography aims to record, on a CCD camera without objective lens,
the intensity distribution of the diffraction pattern of an object
illuminated by a monochromatic continuous wave~\cite{Nicolas_2}. The
numerical and experimental set-up is represented in Fig.
(\ref{fig1}). The incident Gaussian beam propagates in free space
over a distance $z_p$ and illuminates the pipe. Here, the pipe is
modeled as two thick lenses. Their thickness is denoted by $e$. The
opaque object is located between these two lenses: \textbf{at a distance $\delta$ from the first thick lens, and at a distance $z_i$ from the second lens.} The CCD sensor is
located at the distance $z$ next the pipe and records the intensity
distribution of the diffraction pattern.

    \subsection{Intensity distribution in the CCD sensor plane}
In this part, we consider the propagation of a Gaussian beam through
our optical system. In the beam waist plane, located at a distance
$-z_p$ from the pipe, the Gaussian beam, denoted $G$, is defined by:
\begin{equation}
G\left(\mu,\nu\right)=\exp\left(-\frac{\mu^2+\nu^2}{w^2}\right),
\label{ptsource}
\end{equation}
where $\omega$ is the waist width and $\left(\mu,\nu\right)$ are the
coordinates in the beam waist plane.

The propagation of the Gaussian beam through the pipe, to the CCD
sensor is decomposed into two steps. The first step is the
propagation of the illuminating beam (\emph{i.e}
$G\left(\mu,\nu\right)$) from the beam waist plane to the front of
the opaque object. Using the generalized Huygens-Fresnel
integral~\cite{Palma, Lambert, Yura}, the amplitude of the field in
the pipe, denoted $G_1$, is defined by:
\begin{multline}
G_{1}\left(\xi,\eta\right)=\frac{\exp\left(i\frac{2\pi}{\lambda}E_1\right)}{i\lambda \sqrt{B^x_{1}B^y_{1}}}\int_{\mathbb{R}^2}G\left(\mu,\nu\right)\exp\left[i\frac{\pi}{\lambda B^x_{1}}\left(A^x_{1}\mu^2 -2\xi \mu + D^x_{1} \xi^2\right)\right]\\
\times \exp \left[i\frac{\pi}{\lambda B^y_1}\left(A^y_1\nu^2 -2\eta
\nu + D^y_1 \eta^2\right)\right]d\mu d\nu, \label{genefresnel}
\end{multline}
$A^{x,y}_1$, $B^{x,y}_1$, $D^{x,y}_1$ are given by $M_1^{x,y}$
matrices of Eq. (\ref{ABCD1}) in App. (\ref{appA}). To obtain the
relation of the intensity distribution, the linear canonical
transformation is used. For the basic definitions and properties of
the linear canonical transformation, we refer to~\cite{Ozaktas}.
This transformation is a parameterized integral operator of
parameters $A$, $B$, $C$ and $D$. Each parameter represents a
coefficient of a transfer matrix denoted $M$ (\emph{Cf} appendix
\ref{appA}). In our case the Fresnel transform is
considered~\cite{Goodman,Siegman}. It should be noted that the
values of the coefficients are different in both $\xi$ and $\eta$
direction. This is due to the cylindrical geometry of the pipe. The
distance $E_1=z_p+e+\delta$ corresponds to the waist-object
distance. Using the same formalism, the amplitude in the CCD plane,
denoted $G_2\left(x,y\right)$, can be written as:
\begin{multline}
G_2\left(x,y\right)=\frac{\exp\left(i\frac{2\pi}{\lambda}E_2\right)}{i\lambda\sqrt{B^x_2B^y_2}}\int_{\mathbb{R}^2}G_1\left(\xi,\eta\right)\left[1-T\left(\xi,\eta\right)\right]\exp\left[i\frac{\pi}{\lambda B^x_2}\left(A^x_2\xi^2-2x\xi+D^x_2x^2\right)\right]\\
\times\exp\left[i\frac{\pi}{\lambda
B^y_2}\left(A^y_2\eta^2-2y\eta+D^y_2y^2\right)\right]d\xi d\eta,
\label{HF_G2}
\end{multline}
where the parameters $A^{x,y}_2$, $B^{x,y}_2$, $D^{x,y}_2$ are given
by $M_2^{x,y}$ matrices of Eq. (\ref{ABCD2}). The distance
$E_2=z_i+e+z$ is the distance between the object and the CCD sensor.

The spatial transmittance of the opaque 2D-object is defined by
$\left[1-T\left(\xi,\eta\right)\right]$. Here
$T\left(\xi,\eta\right)$ can be expressed as a superposition of
Gaussian functions~\cite{Wen1987,Wen1988}, such as:
\begin{equation}
T\left(r\right)=\sum_{k=1}^NA_k\exp\left(-r^TR^TP_kRr\right),
\label{aperture1}
\end{equation}
with
        \begin{equation}
            R=\left(%
\begin{array}{cc}
  \cos\theta & \sin\theta \\
  -\sin\theta & \cos\theta \\
\end{array}%
\right),
    \end{equation}
and
        \begin{equation}
            P_k=\left(%
\begin{array}{cc}
  \frac{B_k}{a^2} & 0 \\
  0 &  \frac{B_k}{b^2}\\
\end{array}%
\right).
    \end{equation}
This expression permits to deal with non symmetrical optical systems
and elliptical opaque objects~\cite{Zheng,Du2006-1,Du2006-2,Du2007}.
This Gaussian decomposition is very convenient; as a matter of fact
it allows us to establish analytical expression of the diffracted
pattern in the CCD plane. \textbf{It also allows us to simulate
holograms of 3D opaque objects whose 2D projection is elliptic or
circular~\cite{Slimani1984} (\emph{e.g.} spheroids, fibers ...).}

\begin{figure}[t]
\centering
\includegraphics*[width=15cm]{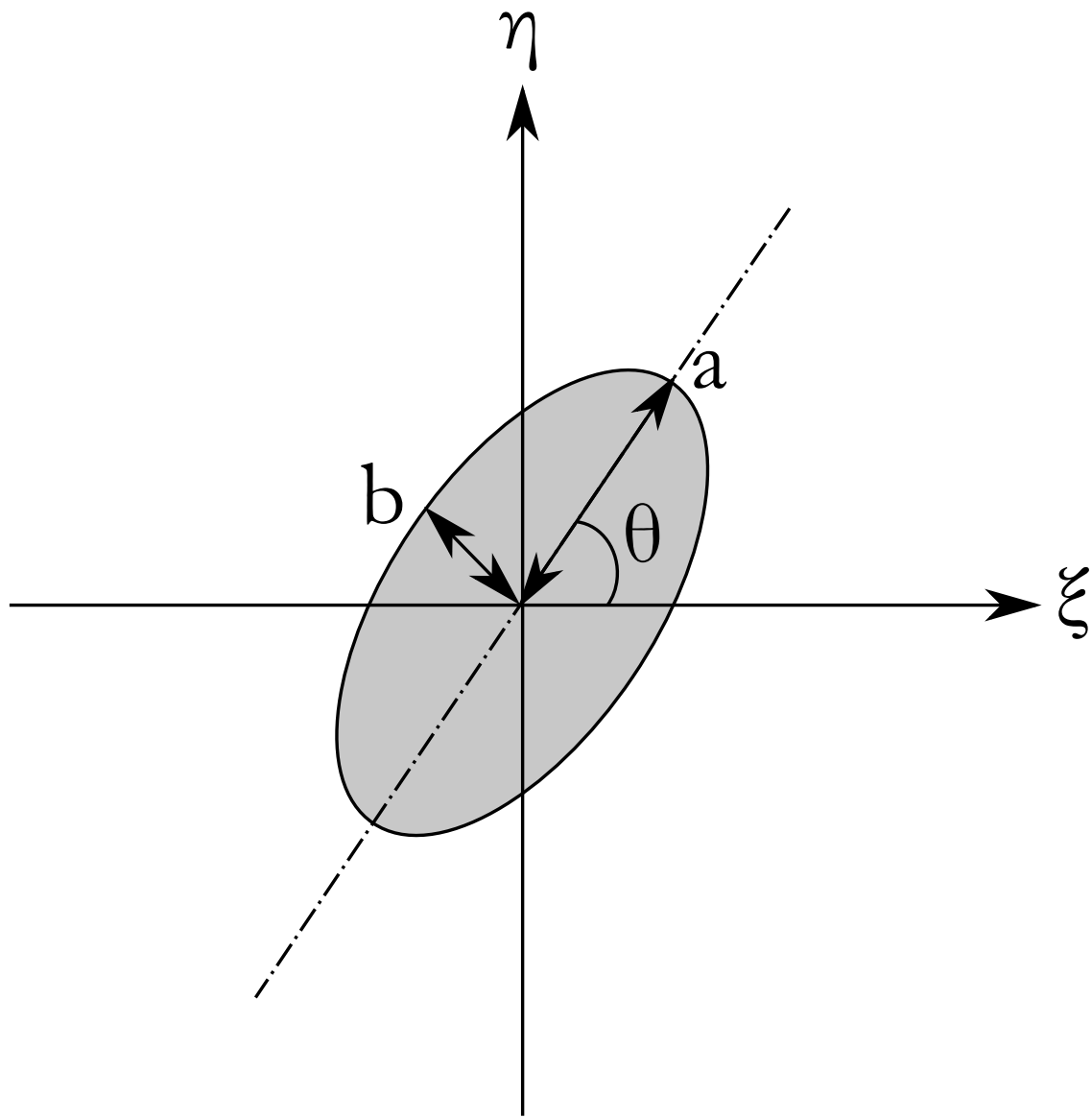}
\caption{Schematic representation of the object.}\label{fig2}
\end{figure}

The object parameters are illustrated on Fig. (\ref{fig2}). The
coefficients $a$ and $b$ are the object radii within $\eta$ and
$\xi$ axis respectively, $\theta$ is the angle between elliptical
aperture principal axis and $\xi$ axis. The $A_k$ and $B_k$
coefficients are determined by numerical resolution of the Kirchhoff
equation~\cite{Wen1988}. Let $R_{ell}=b/a$ representing the particle
ellipticity. Considering $R_{ell}=1$ lead to the simulation of a
circular particle, whereas $R_{ell}\neq 1$ is used to simulate
elliptical particles. The particular case of the opaque fiber,
\textbf{parallel to the pipe axis}, is obtained when
$R_{ell}\rightarrow 0$.

In further developments, $\theta$ will be considered to be $\pi/2$.
With this assumption $T\left(\xi,\eta\right)$ becomes:
\begin{equation}
T\left(\xi,\eta\right)=\sum_{k=1}^{N}A_k\exp\left[-\frac{B_k}{b^2}\left(\xi^2+R_{ell}^2\eta^2\right)\right].
\label{T}
\end{equation}

To simulate our opaque objects, $N$ is fixed to 10. From Eq.
(\ref{HF_G2}), $G_2\left(x,y\right)$ is split into two integrals
denoted $R\left(x,y\right)$ for the {\bf r}eference beam and
$O\left(x,y\right)$ for the {\bf o}bject beam so that :
\begin{equation}
G_2\left(x,y\right)=\frac{\exp\left(i\frac{2\pi}{\lambda}E_2\right)}{i
\lambda
\sqrt{B^x_2B^y_2}}\left[R\left(x,y\right)-O\left(x,y\right)\right],
\end{equation}
with
\begin{multline}
R\left(x,y\right)=\int_{\mathbb{R}^2}G_1\left(\xi,\eta\right)\exp\left[i\frac{\pi}{\lambda B^x_2}\left(A^x_2\xi^2-2x\xi+D^x_2x^2\right)\right]\\
\times\exp\left[i\frac{\pi}{\lambda
B^x_2}\left(A^y_2\eta^2-2y\eta+D^y_2y^2\right)\right]d\xi d\eta,
\label{HF_I1}
\end{multline}
and
\begin{multline}
O\left(x,y\right)=\int_{\mathbb{R}^2}G_1\left(\xi,\eta\right)T\left(\xi,\eta\right)\exp\left[i\frac{\pi}{\lambda B^x_2}\left(A^x_2\xi^2-2x\xi+D^x_2x^2\right)\right]\\
\times\exp\left[i\frac{\pi}{\lambda
B^y_2}\left(A^y_2\eta^2-2y\eta+D^y_2y^2\right)\right]d\xi d\eta.
\label{HF_I2}
\end{multline}

The functions $R\left(x,y\right)$ and $O\left(x,y\right)$ are
respectively given by:
\begin{multline}
R\left(x,y\right)=\frac{\exp\left(i\frac{2\pi}{\lambda}E_1\right)}{i\lambda\sqrt{B^x_1B^y_1}}K_1^xK_1^yK_2^xK_2^y\\
\times\exp\left[-\frac{\pi}{\lambda}\left(\frac{N_x}{B^x_2}x^2+\frac{N_y}{B^y_2}y^2\right)\right]\exp\left[i\frac{\pi}{\lambda}\left(\frac{M_x}{B^x_2}x^2+\frac{M_y}{B^y_2}y^2\right)\right],
\label{I1}
\end{multline}
and
\begin{multline}
O\left(x,y\right)=\frac{\exp\left(i\frac{2\pi}{\lambda}E_1\right)}{i\lambda\sqrt{B^x_1B^y_1}}K_1^xK_1^y\exp\left[i\frac{\pi}{\lambda}\left(\frac{D^x_2}{B^x_2}x^2+\frac{D^y_2}{B^y_2}y^2\right)\right]\sum_{k=1}^{N}A_kK_2^{{x}_{eq}}K_2^{{y}_{eq}}\\
\times\exp\left[-\frac{\pi}{\lambda}\left(\frac{N_{x_{eq}}}{B^x_2}x^2+\frac{N_{y_{eq}}}{B^y_2}y^2\right)\right]\exp\left[i\frac{\pi}{\lambda}\left(\frac{M_{x_{eq}}}{B^x_2}x^2+\frac{M_{y_{eq}}}{B^y_2}y^2\right)\right].
\label{I2}
\end{multline}
Values of the different parameters of Eqs. (\ref{I1}) and (\ref{I2})
are defined in App. (\ref{appB}).

After theoretical developments the intensity distribution, denoted
$I\left(x,y\right)$, recorded by the CCD sensor is:
\begin{equation}
I\left(x,y\right)=G_2\left(x,y\right)\overline{G_2\left(x,y\right)}=\frac{1}{\lambda^2B^x_2B^y_2}\left(\left|R\right|^2-2\Re\left\{R\overline{O}\right\}+\left|O\right|^2\right),
\label{Intensity}
\end{equation}
where the upper bar denotes the complex conjugate and $\Re$ the real
part. The square modulus $\left|R\right|^2$ corresponds to the
directly transmitted beam whereas $\left|O\right|^2$ is associated
with the diffracted part of the beam.

To illustrate Eq. (\ref{Intensity}), the optical set-up given on
Fig. (\ref{fig1}) is considered. We investigate the diffraction
pattern obtained with an opaque fiber \textbf{parallel to the pipe
axis}. From the Gaussian decomposition of the object function,
opaque fibers, parallel to the pipe axis, can be obtained taking
$R_{ell}\rightarrow 0$. The value of the opaque fiber diameter is
$2b=51.8\;µm$ and $a\rightarrow\infty$. The glass-made ($n_1=1.5$)
pipe is filled with water of refractive index $n_2=1.33$. The beam
propagates in free space over $z_p=325\;mm$. The dimensions $\delta$
and $z$ are fixed to 18 $mm$ and 23 $mm$ respectively.

\begin{figure}[t]
\centering
\includegraphics*[width=15cm]{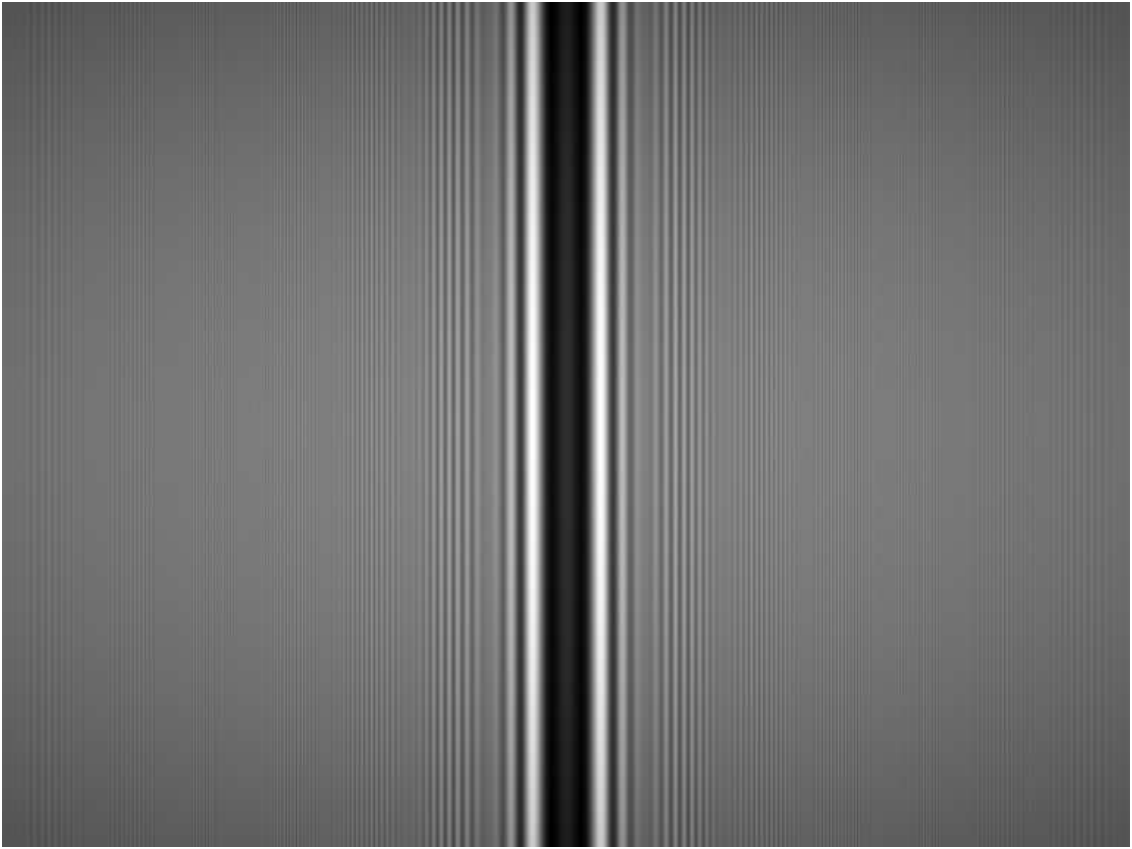}
\caption{Simulation of the diffraction pattern of a 51.8 $\mu m$
opaque fiber, parallel to the axis of a glass pipe, $n_2=1.33$,
$\lambda=632.8nm$, $z=23mm$, $\delta=18mm$.}\label{fig3}
\end{figure}

\begin{figure}[t]
\centering
\includegraphics*[width=15cm]{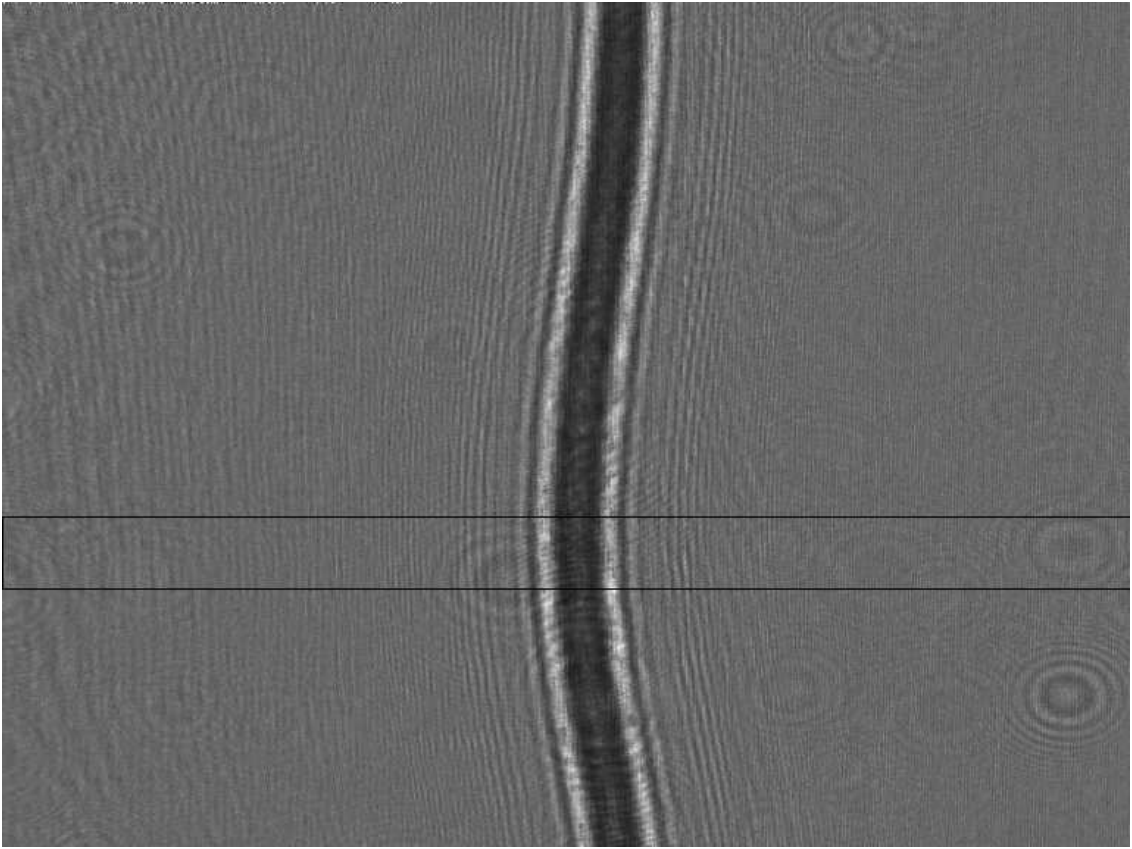}
\caption{Experimental diffraction pattern of a 51.8 $\mu m$ opaque
fiber, parallel to the axis of a glass pipe, $n_2=1.33$,
$\lambda=632.8nm$, $z=23mm$, $\delta=18mm$.}\label{fig4}
\end{figure}

\begin{figure}[t]
\centering
\includegraphics*[width=15cm]{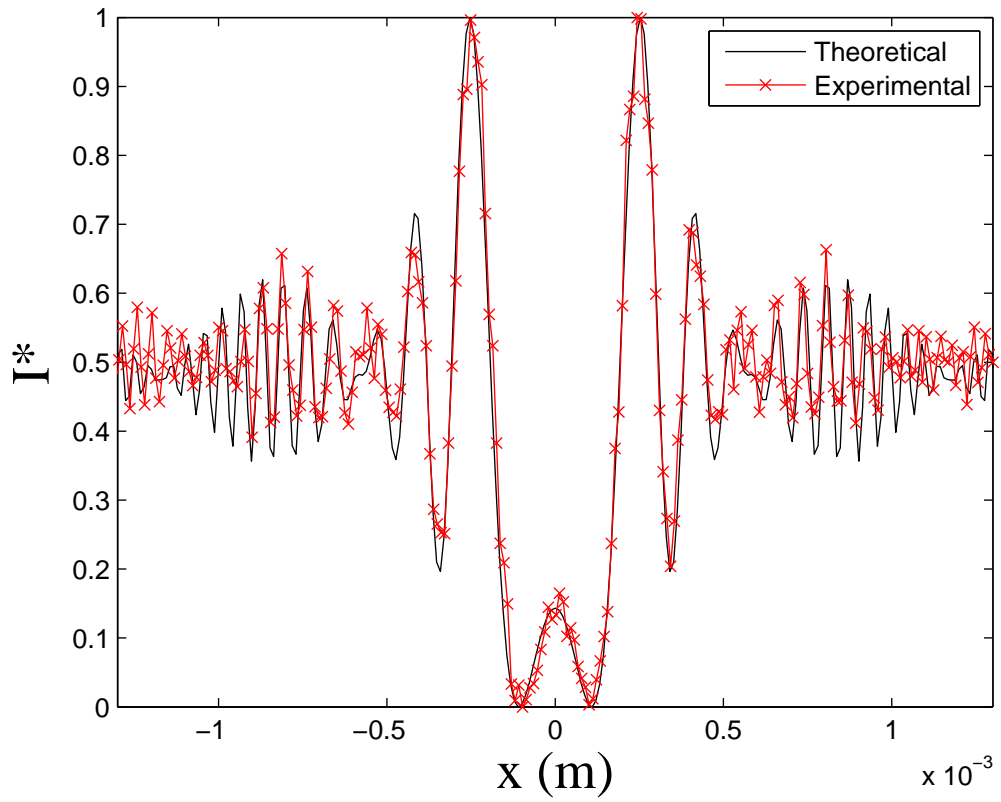}
\caption{Comparison between simulated and experimental intensity
distributions}\label{fig5}
\end{figure}

The image of Fig. (\ref{fig3}) illustrates the intensity
distribution of the diffracted field. The simulation is carried out
by calculating the gray level of each pixel using Eq.
(\ref{Intensity}). The size of the hologram is $768\times576$
pixels, and pixel pitch is $11\;µm$. To validate our simulation, an
experiment, which consists in placing an opaque fiber in the pipe,
was performed. The hologram of Fig (\ref{fig4}) represents the
intensity of the diffraction recorded with the CCD sensor using the
same parameters than previously. This illustration reveals a good
accordance between numerical and experimental diffraction patterns.
To confirm this point, the transverse intensity profiles obtained
from Fig. (\ref{fig3}) and Fig. (\ref{fig4}) are presented on Fig
(\ref{fig5}). Here the normalized intensity $I^*$ is plotted against
the x-axis. The intensity profiles are calculated by cumulating the
gray levels along the $\eta$-axis (direction of the fiber) over 50
rows, \textbf{which are figured out by the rectangular selection on
Fig. (\ref{fig4}).} \textbf{The diffraction pattern of Fig.
(\ref{fig4}) was shifted by $75\;µm$ along x-axis so that numerical
and experimental results can be compared.} This figure shows the
good agreement between numerical and experimental data.

\subsection{Case of thin lenses}
We have proposed a theoretical model allowing us to deal with
propagation of a laser beam through a thick pipe. In a former study,
the effect of cylindrical lenses on the diffraction pattern of a
particle has been investigated in details, leading to the expression
of the intensity distribution in the CCD plane~\cite{Nicolas}. Using
this approach we have a great opportunity to confirm our pipe model
results.

In the following developments we aim to compare our thick and
cylindrical pipe model with a thin cylindrical lenses approach. The
phase transformation due to a thin lens is~\cite{Goodman,Nicolas}:
\begin{equation}
\Phi\left(x_{l},y_{l}\right)=\exp\left[-i\frac{\pi}{\lambda}\left(\frac{x_l^2}{f_x}+\frac{y_l^2}{f_y}\right)\right].
\end{equation}
Here $\left(x_l,y_l\right)$ are the coordinates in the lens plane
and $f_{x,y}$ are the focal lengths of the lens in both directions.
Using Huygens-Fresnel integral~\cite{Goodman} the intensity
distribution in the sensor plane is:
\begin{equation}
I_{thin}\left(x,y\right)=\left|\frac{\exp\left(i\frac{2\pi}{\lambda}E_2\right)}{i\lambda
E_2}\left[R_{thin}\left(x,y\right)-O_{thin}\left(x,y\right)\right]\right|^2.
\label{I_thin}
\end{equation}
Here, $R_{thin}$ and $O_{thin}$ can be expressed in a quite similar
form than amplitude distributions given in Eqs. (\ref{I1}) and
(\ref{I2}).

\begin{figure}[t]
\centering
\includegraphics*[width=15cm]{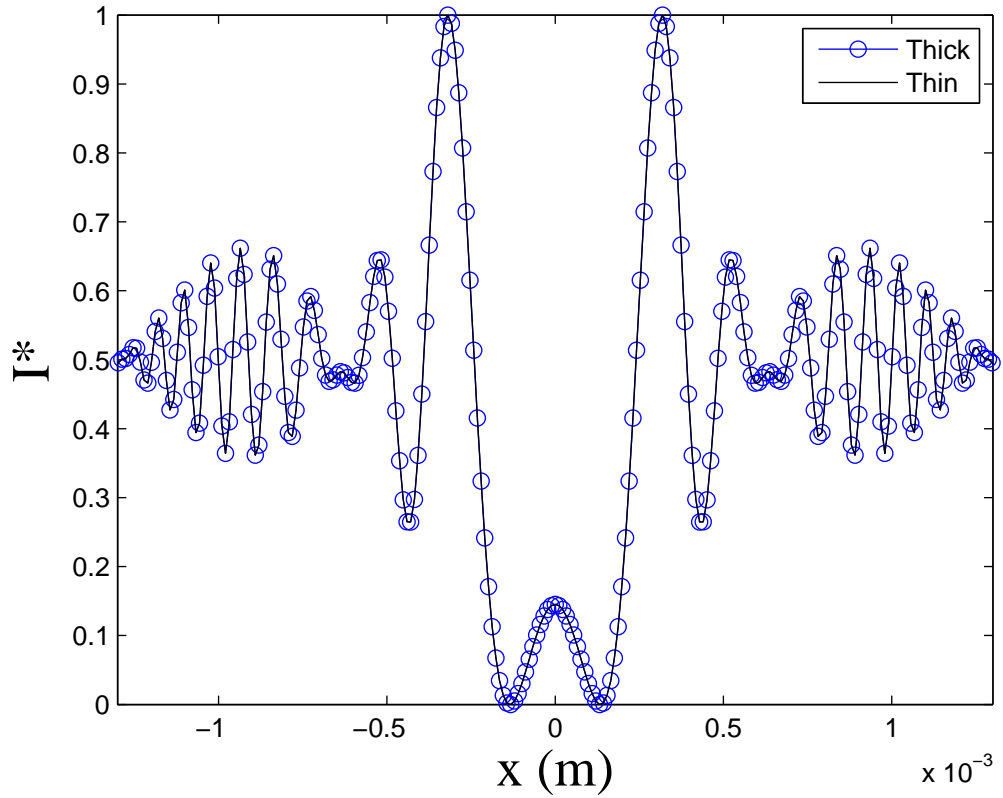}
\caption{Normalized intensity distribution recorded by the CCD:
comparison between thin lens and thick lens models, with $e=0\;mm$.
Simulation parameters are : $2b=51.8\;µm$, $a\rightarrow\infty$,
$z_p=325\ mm$, $n_2=1$, $\delta=18\ mm$ and $z=23\ mm$}\label{fig6}
\end{figure}

The transverse intensity distribution profile presented in Fig.
(\ref{fig6}) shows:
\begin{equation}
\lim_{e\rightarrow 0}I\left(x,y\right)=I_{thin}\left(x,y\right),
\end{equation}
\textbf{remembering that $e$ represents the glass thickness. This
term is contained in matrices $M_1^{x,y}$ and $M_2^{x,y}$ (see
Eqs. (\ref{ABCD1}) and (\ref{ABCD2})).}

For this example: $2b=51.8\;µm$, $a\rightarrow\infty$, $z_p=325\
mm$, $n_2=1$, $\delta=18\ mm$ and $z=23\ mm$. The comparison with
other parameter values leads to the same conclusion. As a result, we
can consider that our results are consistent with those of Ref.
\cite{Nicolas} and that our model is versatile enough to deal with
various pipe shapes and thickness.

In this section a numerical model allowing to treat thick optical
systems has been presented. We now aim to reconstruct the image of
the object by means of FRFT from the calculated intensity
distribution of the diffraction pattern (Eq. (\ref{Intensity})). In
section \ref{Sec3}, mathematical definition of the FRFT is recalled
leading to the reconstruction of holograms recorded in pipe systems.

\section{Fractional Fourier transformation analysis of in-line holograms}
\label{Sec3}

\subsection{ Two-dimensional Fractional Fourier transformation }

The FRFT is a generalization of the classical Fourier transform.
This integral operator has numerous applications in signal
processing~\cite{Ozaktas}. Its mathematical expression is the
following~\cite{MBride1987,Namias1980,Lohmann1993}: the FRFT of
order $a_x=(2 \alpha_x)/\pi$ and $a_y=(2 \alpha_y)/\pi$ (for $x$ and
$y$ cross section respectively), with $0\leq \left| \alpha_x \right|
\leq\pi/2$ and $0\leq \left| \alpha_y \right|\leq \pi/2$, of a two
dimensional function $I\left(x,y\right)$ is
\begin{equation}\label{FRFT:equ1}
\mathscr{F}_{\alpha_{x},\alpha_{y}}[I(x,y)](x_{a},y_{a})=\int_{\mathbb{R}^{2}}
N_{\alpha_{x}}(x,x_{a})\,N_{\alpha_{y}}(y,y_{a})I(x,y)\,dx\,dy,
\end{equation}
where the kernel of the fractional operator is defined by
\begin{equation}\label{FRFT:equ2}
N_{\alpha_{p}}(x,x_{a})=C(\alpha_{p})\exp\left(i\pi\frac{x^{2}+x_{a}^{2}}{s_p^{2}\tan
\alpha_{p}}\right) \exp\left(-\frac{i2\pi x_{a}
x}{s_p^{2}\sin\alpha_{p}}\right),
\end{equation}
and
\begin{equation}\label{FRFT:equ3}
C(\alpha_{p})=\frac{\exp(-i(\frac{\pi}{4}\textrm{sign}(\sin\alpha_{p})-\frac{\alpha_{p}}{2}))}
{|s_p^{2}\sin\alpha_{p}|^{1/2}}\,.
\end{equation}
Here $p=x,y$. Generally, the parameter $s_p$ is considered as a
normalization constant. In our case, its value is defined from the
experimental set-up according to~\cite{Mas}:
$s_p^2=N_p^{px}\cdot\delta_p^2$. The number of samples is $N_p^{px}$
in both intensity distribution $I\left(x,y\right)$ and fractional
domain. $\delta_p$ is the sampling period along the two axes of the
image. $C(\alpha_{p})$, which is a function of the fractional order,
insures the energy conservation law to be valid.
\textbf{Discretization of the FRFT kernel is performed using an
orthogonal projection method proposed by Pei \emph{et
al.}~\cite{Pei1999}.}

\subsection{Optimal fractional orders to refocus numerically over the object}

In our case, the numerical reconstruction can be considered as a
numerical refocusing over the object~\cite{Verrier}. To do that, the
quadratic phase, denoted $\varphi=\arg\left(R\overline{O}\right)$,
and contained in the term $2\Re\left\{R\overline{O}\right\}$ of Eq.
(\ref{Intensity}) must be evaluated. This term is composed of a
linear chirp modulated by a sum of complex Gaussian functions.
Information about the distance between the CCD sensor and the object
is carried by the linear chirp ($B_{2_{x,y}}$) whereas information
about object size is carried by the modulation ($M_{{x,y}_{eq}}$ and
$N_{{x,y}_{eq}}$). The optimal reconstruction of the image consists
in compensating the quadratic phase terms of the intensity
distribution~\cite{Nicolas}.

The quadratic phase can be determined from Eqs. (\ref{I1}) and
(\ref{I2})
\begin{equation}
\varphi=\frac{\pi}{\lambda}\left[\left(\frac{M_x-D^x_2}{B^x_2}\right)x^2+\left(\frac{M_y-D^y_2}{B^y_2}\right)y^2\right],
\label{phi}
\end{equation}
thus $\Re\left\{R\overline{O}\right\}$ can be written as:
\begin{equation}
\Re\left\{R\overline{O}\right\}=\left|R\overline{O}\right|\cos\left(i\varphi\right).
\end{equation}

The quadratic phase term contained in the FRFT kernel, denoted
$\varphi_{a}$, is given by
\begin{equation}
\varphi_{a}=\pi\left(\frac{\cot \alpha_x}{s_x^2}x^2+\frac{\cot
\alpha_y}{s_y^2}y^2\right). \label{phia}
\end{equation}

The image reconstruction is obtained by applying the FRFT to the
intensity distribution of Eq. (\ref{Intensity}):
\begin{equation}
\mathscr{F}_{\alpha_{x},\alpha_{y}}[I(x,y)]\propto\mathscr{F}_{\alpha_{x},\alpha_{y}}[\left|R\right|^2+\left|O\right|^2]-2\mathscr{F}_{\alpha_{x},\alpha_{y}}[\left|R\overline{O}\right|\cos\varphi].
\label{FRFT_I}
\end{equation}
The terms $\left|R\right|^2$ and $\left|O\right|^2$ contain no
linear chirps, thus they do not influence the optimal fractional
orders of reconstruction to be determined. Only the second term is
useful for image reconstruction. By noting that
$2\cos\varphi=\exp\left(-i\varphi\right)+\exp\left(i\varphi\right)$,
Eq. (\ref{FRFT_I}) becomes:
\begin{multline}
\mathscr{F}_{\alpha_{x},\alpha_{y}}[I(x,y)]\propto\mathscr{F}_{\alpha_{x},\alpha_{y}}[\left|R\right|^2+\left|O\right|^2]\\
-C\left(\alpha_x\right)C\left(\alpha_y\right)\int_{\mathbb{R}^2}\left|R\overline{O}\right|\exp\left[i\left(\varphi_a-\varphi\right)\right]\exp\left[-i2\pi\left(\frac{x_ax}{s_x^2\sin\alpha_x}+\frac{y_ay}{s_y^2\sin\alpha_y}\right)\right]dxdy\\
-C\left(\alpha_x\right)C\left(\alpha_y\right)\int_{\mathbb{R}^2}\left|R\overline{O}\right|\exp\left[i\left(\varphi_a+\varphi\right)\right]\exp\left[-i2\pi\left(\frac{x_ax}{s_x^2\sin\alpha_x}+\frac{y_ay}{s_y^2\sin\alpha_y}\right)\right]dxdy.
\end{multline}
The best hologram reconstruction is reached when one of the
quadratic phase term is brought to zero. Thus reconstruction is
performed if
\begin{equation}
\varphi_a\pm\varphi=0. \label{cond}
\end{equation}
The optimal fractional orders $\alpha_{x}^{opt}$ and
$\alpha_{y}^{opt}$ are defined from Eqs. (\ref{phi}), (\ref{phia})
and (\ref{cond}) and take the values:
\begin{equation}
\alpha_x^{opt}=\arctan\left[\mp\frac{B^x_2\lambda}{s_x^2\left(M_x-D^x_2\right)}\right],\quad\alpha_y^{opt}=\arctan\left[\mp\frac{B^y_2\lambda}{s_y^2\left(M_y-D^y_2\right)}\right].
\label{orders}
\end{equation}
It should be noted that we have checked that if we consider
$M_{2_{x,y}}$ to be free space propagation, leading to
$B^{x,y}_2=z_i+e+z$ and $D^{x,y}_2=1$, the expression of the optimal
fractional order of Ref. ~\cite{Nicolas} are recovered.

\subsection{Numerical simulations}
\begin{figure}[t]
\centering
\includegraphics*[width=15cm]{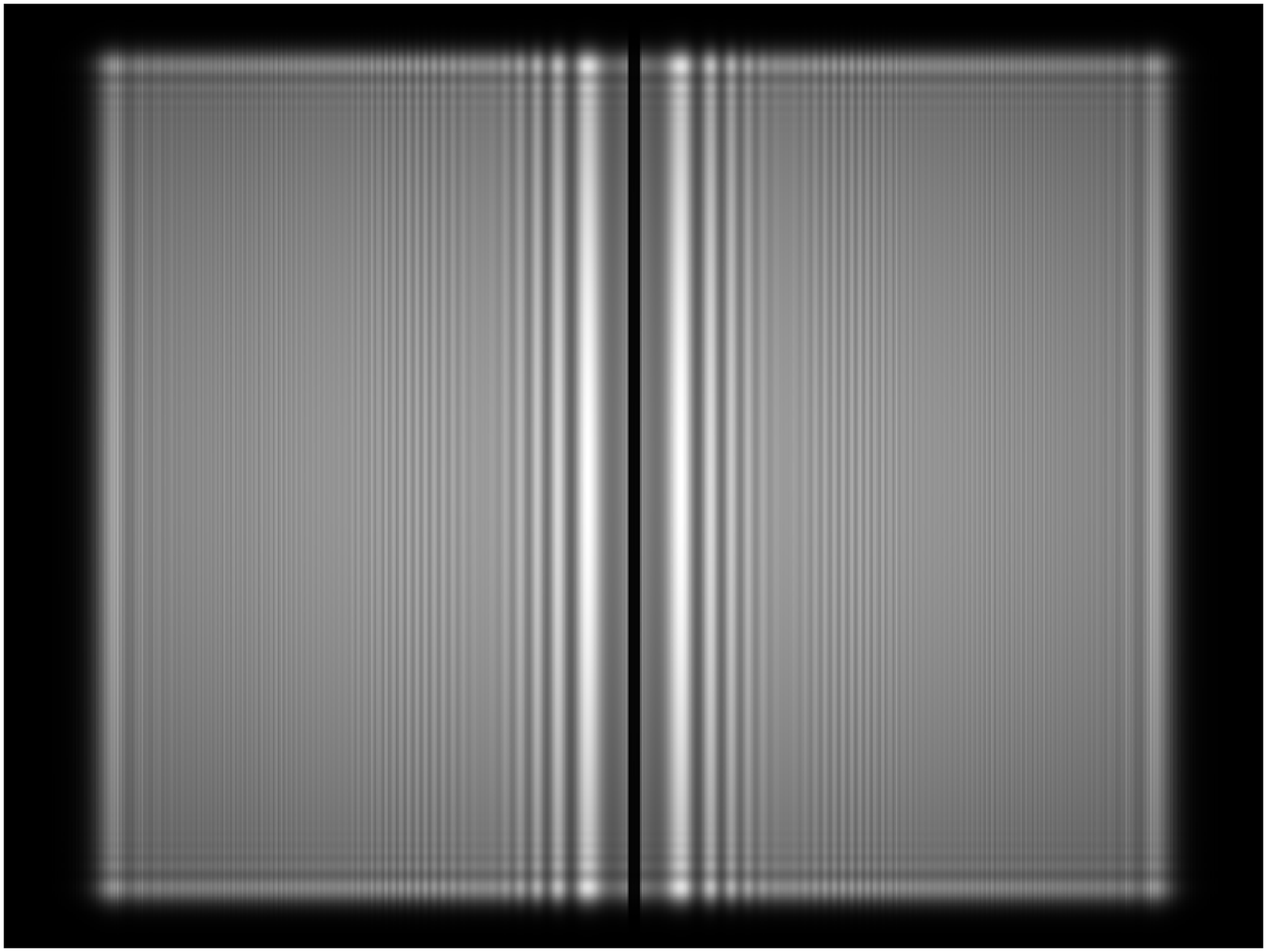}
\caption{Reconstruction of the fiber image from the diffraction
pattern of Fig. (\ref{fig3}) by FRFT with $a_x^{opt}=0.337$ and
$a_y^{opt}=0.273$.}\label{fig7}
\end{figure}

Experimental set-up of Fig. (\ref{fig1}) is used to perform
simulations. An application of the model is illustrated on Fig.
(\ref{fig3}).  Recall that we consider an opaque fiber ($R_{ell}
\rightarrow 0$) $2b=51.8\mu m$ in width. The glass made pipe is
filled with water ($n_2=1.33$). $\delta$ and $z$ are set to 18 $mm$
and 23 $mm$ respectively. $Rx_1, Ry_1, Ry_2\rightarrow\infty$ and
$Rx_2=18\ mm$ are the first thick lens curvatures along $x$- and
$y$-axis. The curvature radii of the second thick lens have the same
values but opposite signs. With these parameters and owing to Eq.
(\ref{orders}) the optimal fractional orders within this
configuration are
\begin{equation}
a_x^{opt}=0.337,\quad a_y^{opt}=0.273. \label{alphath}
\end{equation}
The reconstruction of the fiber with FRFT can be seen on Fig.
(\ref{fig7}). As we can see on this figure, the reconstructed image
is disturbed by background fringes. This effect is due to the
in-line configuration and is commonly known as the twin image
effect~\cite{Gabor}.

\subsection{Experimental results}

Since now, we have presented a method to simulate holograms in thick
optical systems. Reconstruction has been successfully performed
thanks to FRFT. In order to validate theoretical developments and
simulations, a glass pipe with curvatures  $Rx_1, Ry_1,
Ry_2\rightarrow\infty$ and $Rx_2=18\ mm$ has been used for the
experiments. Curvatures on the other side of the pipe are deduced
from $Rx_1, Ry_1, Rx_2, Ry_2$ by taking negative values.

\begin{figure}[t]
\centering
\includegraphics*[width=15cm]{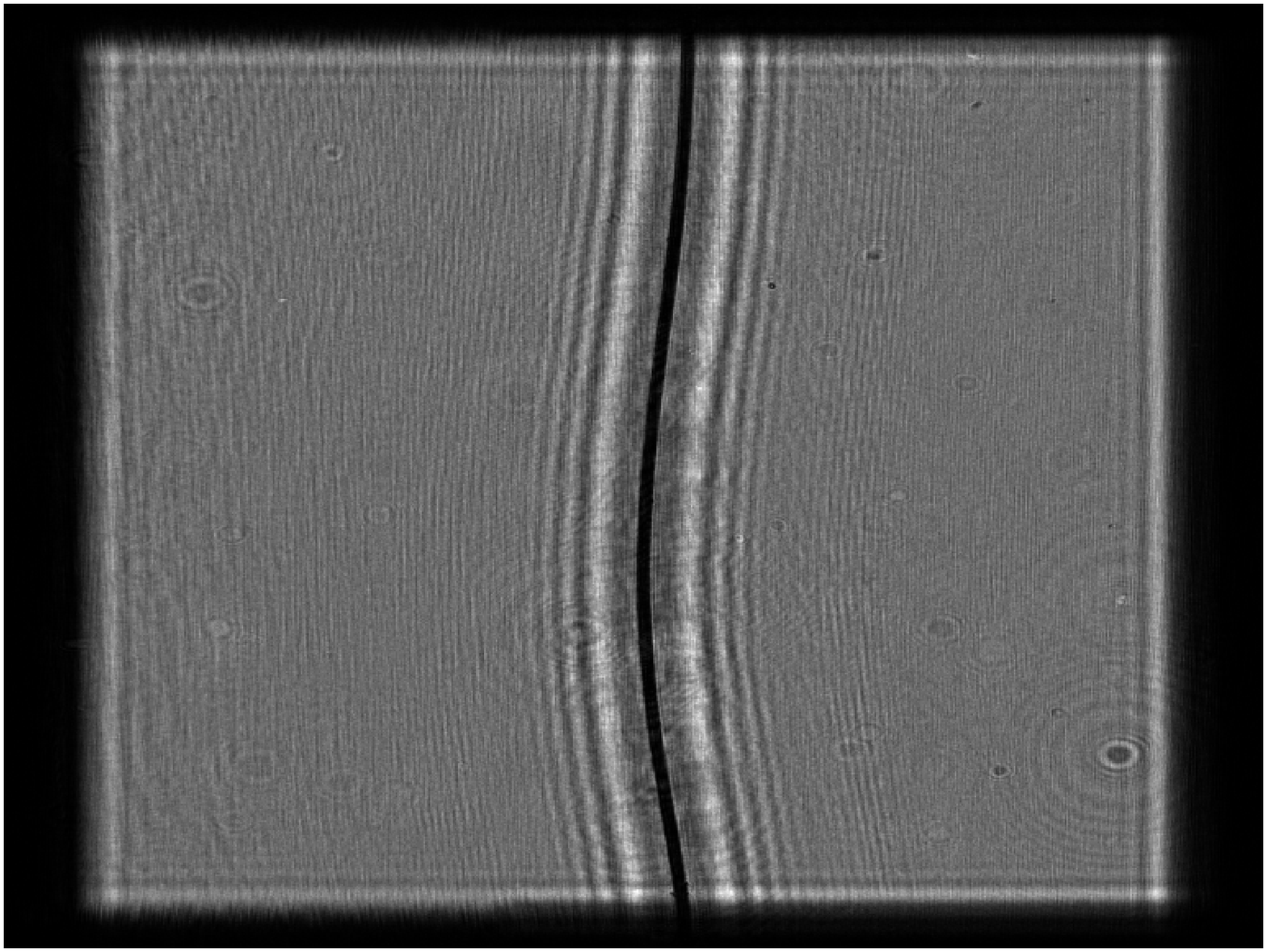}
\caption{Reconstruction of the fiber image from the diffraction
pattern of Fig. (\ref{fig4}) by FRFT with $a_x^{opt}=0.33$ and
$a_y^{opt}=0.27$.}\label{fig8}
\end{figure}

The image of Fig. (\ref{fig4}) represents the intensity of the
diffraction pattern recorded with a $768\times 576\ px^2$ CCD sensor
with $11\ \mu m$ pitch. The object is a $2b=51.8 \mu m$ opaque
fiber, $\delta$ and $z$ are approximately and respectively equal to
$18\ mm$ and $23\ mm$. Theoretical orders associated with this
experiment are given by Eq. (\ref{alphath}). To perform FRFT on this
experimental image, FRFT orders have been adjusted to obtain the
best image of the fiber \emph{i. e} the best contrast between the
reconstructed fiber and the background. Thus doing, the accuracy on
the fractional order is approximately $10^{-2}$. Reconstruction is
presented on Fig. (\ref{fig8}), estimated optimal fractional orders
are $a_x=0.33$ and $a_y=0.27$. These values are very close to the
theoretical ones.

\begin{figure}[t]
\centering
\includegraphics*[width=15cm]{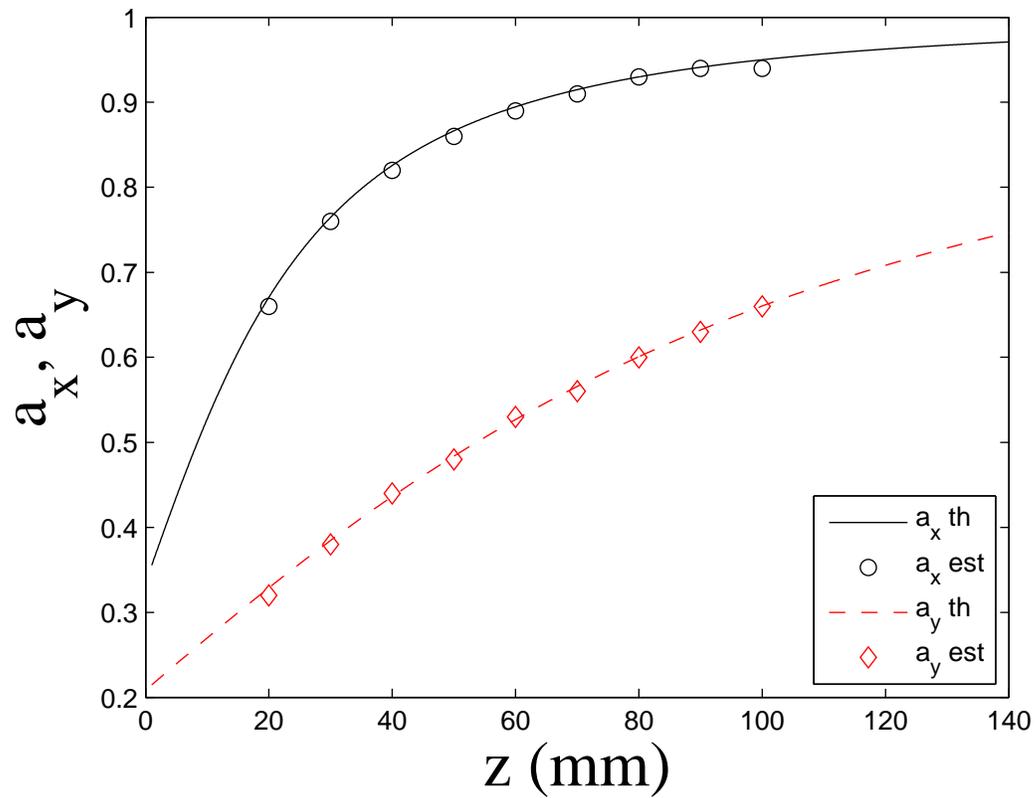}
\caption{Comparison between theoretical fractional orders and
optimal fractional orders estimated from the experimental
holograms.}\label{fig9}
\end{figure}

This good accordance is confirmed by the curves of Fig.
(\ref{fig9}). Here theoretical values of $a_x^{opt}$ (solid line)
and $a_y^{opt}$ (dashed line) are plotted versus the distance
between the pipe and the CCD sensor. Theoretical values corresponds
to an air filled glass pipe, $\delta$ is fixed to $10\ mm$ and $z$
varies from 0 to $140\ mm$. Note that, when $z=0$, $a_{x,y}\neq 0$,
this is due to the fact that, in this case, $z$ represents the
distance between the pipe and the CCD sensor instead of the distance
between the object and the CCD sensor which is considered in
\cite{Nicolas}. Experiments performed with this configuration allow
us to plot estimated values of $a_x^{opt}$ (circles) and $a_y^{opt}$
(diamonds) on the same graph. As shown by Fig. (\ref{fig9}),
estimated and theoretical values are closely linked. Theoretical
representation hereby presented is therefore well adapted to these
optical systems.

\begin{figure}[t]
\centering
\includegraphics*[width=15cm]{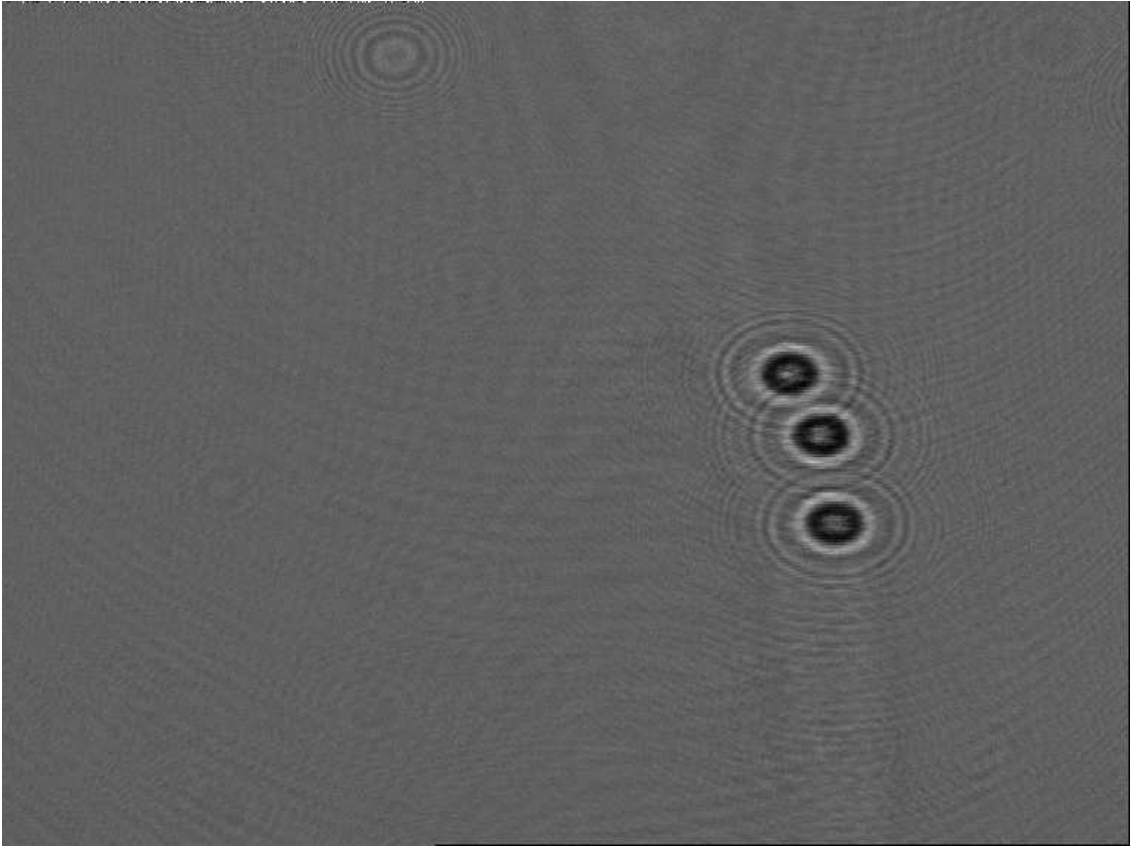}
\caption{Experimental diffraction pattern of $100 µm$ latex beads.
$z_p=325\ mm$, $z=23\ mm$.}\label{fig10}
\end{figure}

\textbf{Until then, we have performed reconstruction with objects for which position is well known.
We now apply our formalism to reconstruct hologram of latex beads. The intensity distribution of the diffracted field is represented on Fig. (\ref{fig10}).
Here, $100\;µm$ in diameter latex beads are used, $z_p=325\;mm$ (\emph{i.e.} propagation distance between the source and the pipe) and $z=23\;mm$ (\emph{i.e.} distance between the pipe and the CCD sensor). This study gives us the opportunity to compare reconstruction using our novel approach with reconstruction using a thin lens approach. As far as every parameter of the experimental set-up is known, excepting $\delta$, the reconstruction process is the following: for $\delta \in \left[0,36\;mm\right]$ we calculate the optimal FRFT orders using Eq. (\ref{orders}). Doing so, we are able to refocus on the particles which are in the pipe.}

\textbf{In Fig. (\ref{fig6}), it is shown that when the pipe thickness $e$ is close to zero, the two approaches are equivalent. Thus, to reconstruct the hologram of Fig. (\ref{fig10}) with the thin lens approach, we only need to use the orders given in Eq. (\ref{orders}) with $e=0$.}

\begin{figure}[t]
\centering
\subfigure[]{\includegraphics*[width=7cm]{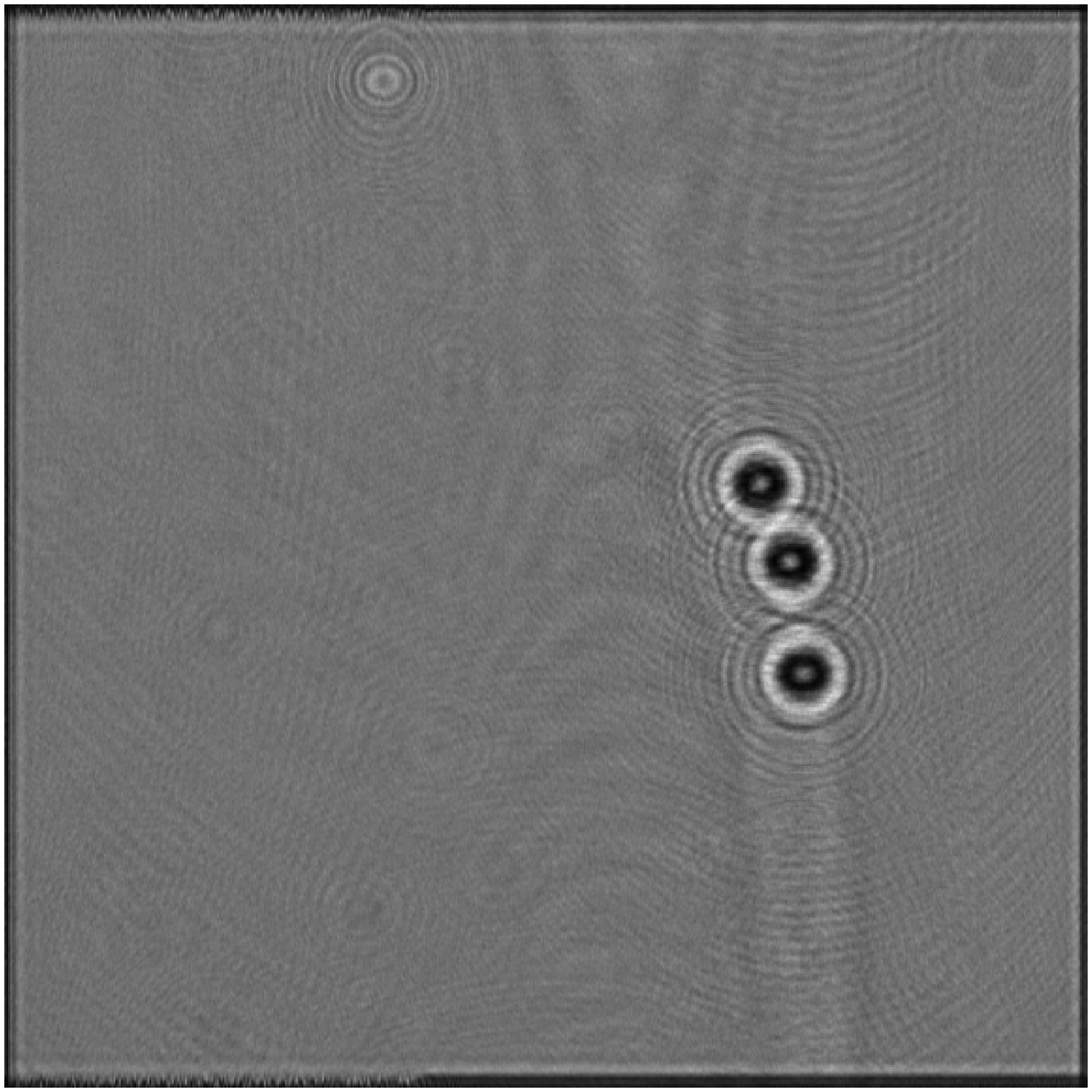}}\quad
\subfigure[]{\includegraphics*[width=7cm]{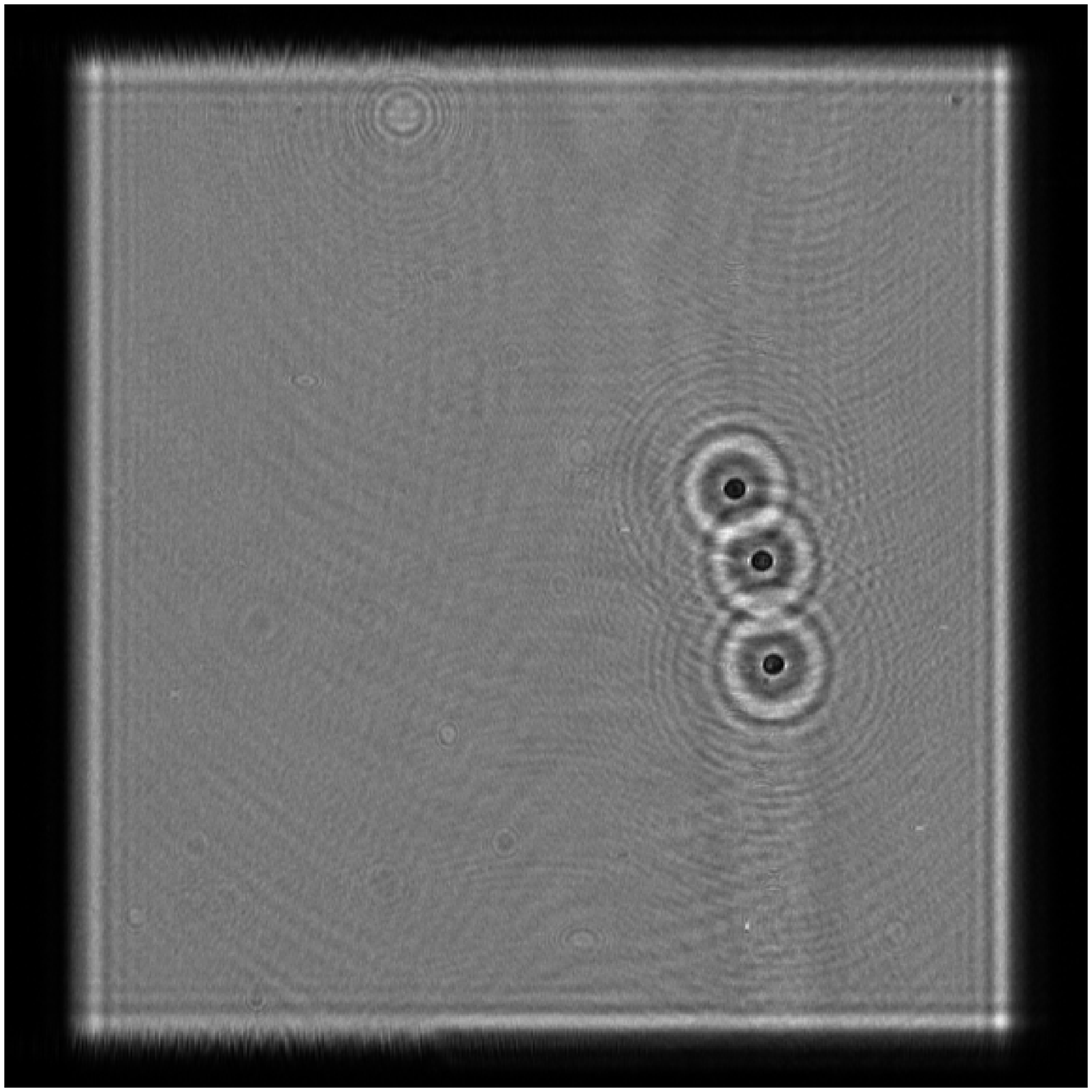}}\quad
\caption{Reconstruction of the latex beads image. (a) Reconstruction
with thin lens parameters. (b) Reconstruction using thick lens approach.}\label{fig11}
\end{figure}

\textbf{Comparison between the two reconstruction methods can be made with Fig. (\ref{fig11}). Here, the reconstruction is realized with the same value of $\delta$ in both cases. We can notice that refocusing on the particle is impossible with the thin lens approach, whereas our thick lens approach allows a good reconstruction of the image of particles.}

\textbf{Therefore, the thick lens formalism is well adapted to pipe flow studies: as we are able to refocus on particles in a pipe, metrologies of particles diameter is possible.}

\section{Conclusion}
In this paper, an analytical expression of the scalar diffraction,
under Fresnel approximations, in thick optical systems such as pipes
is derived. As such optical systems reveals astigmatism, FRFT is a
reliable tool to perform hologram reconstruction. Optimal fractional
orders have been calculated leading to a satisfactory reconstruction
of either numerical or experimental images.  Numerical experiments
have been performed showing a good agreement with experimental
results. \textbf{Comparison between thin lens and thick lens approaches has been performed
showing that the thin lens formalism is not well adapted to the study of pipe flows.}

\section*{Acknowledgments}
These works come within the scope of Hydro-Testing Alliance (HTA)
with the support of the European Commission's Sixth Framework
Programme under DG Research.
\newpage
\appendix

\section{Appendix A : Transfer matrices of the optical system}\label{appA}
Each part of an optical system can be represented by a matrix. Using
this principle, we can build a set of matrices corresponding to our
experiments (Fig. (\ref{fig1})). After the beam shaping step, the
beam propagates in free space over a distance $z_p$. The associated
$M_{z_p}$ matrix is:
    \begin{equation}
    M_{z_p}=\left(%
\begin{array}{cc}
  1 & z_p \\
  0 & 1 \\
\end{array}%
\right). \label{zp}
    \end{equation}

Then the beam is refracted at the interface between free space and
glass ($n_1=1.5$). The curvatures along both directions are $Rx_1$
and $Ry_1$. $M_{Rx_1},\ M_{Ry_1}$ matrices along $x$ and $y$ axis
are:
    \begin{equation}
    M_{Rx_1}=\left(%
\begin{array}{cc}
  1 & 0 \\
  \frac{n_0-n_1}{Rx_1} & 1 \\
\end{array}%
\right),\quad M_{R1_y}=\left(\begin{array}{cc}
  1 & 0 \\
  \frac{n_0-n_1}{Ry_1} & 1 \\
\end{array}%
\right). \label{L1ref1}
    \end{equation}

After refraction, the beam propagates in glass ($n_1$=1.5). For this
step:
        \begin{equation}
    M_{e}=\left(%
\begin{array}{cc}
  1 & \frac{e}{n_1} \\
  0 & 1 \\
\end{array}%
\right), \label{L1prop}
    \end{equation}

Next step is the refraction of the beam at the interface between
glass and the medium inside the pipe (refractive index $n_2$).
Curvature are given by $Rx_2$ and $Ry_2$ in $x$ and $y$ direction.
    \begin{equation}
    M_{Rx_2}=\left(%
\begin{array}{cc}
  1 & 0 \\
  \frac{n_1-n_2}{Rx_2} & 1 \\
\end{array}%
\right),\quad M_{Ry_2}=\left(\begin{array}{cc}
  1 & 0 \\
  \frac{n_1-n_2}{Ry_2} & 1 \\
\end{array}%
\right), \label{L1ref2}
    \end{equation}

We are now in the pipe. To reach the object, we have to propagate
over $\delta$.
        \begin{equation}
    M_{\delta}=\left(%
\begin{array}{cc}
  1 & \frac{\delta}{n_2} \\
  0 & 1 \\
\end{array}%
\right).
    \end{equation}

Doing the same over $z_i$ permit to reach the output of the pipe:
        \begin{equation}
    M_{z_i}=\left(%
\begin{array}{cc}
  1 & \frac{z_i}{n_2} \\
  0 & 1 \\
\end{array}%
\right).
    \label{zi}
    \end{equation}

After a refraction at the interface (curvature $Rx_3$ and $Ry_3$)
between the medium with refractive index $n_2$ and glass:
    \begin{equation}
    M_{Rx_3}=\left(%
\begin{array}{cc}
  1 & 0 \\
  \frac{n_2-n_1}{Rx_3} & 1 \\
\end{array}%
\right),\quad M_{Ry_3}=\left(\begin{array}{cc}
  1 & 0 \\
  \frac{n_2-n_1}{Ry_3} & 1 \\
\end{array}%
\right), \label{L2ref1}
    \end{equation}

a propagation in glass ($n_1$=1.5)
        \begin{equation}
    M_{e}=\left(%
\begin{array}{cc}
  1 & \frac{e}{n_1} \\
  0 & 1 \\
\end{array}%
\right), \label{L2prop}
    \end{equation}

a refraction at the interface (curvature $Rx_4$ and $Ry_4$) between
glass ($n_1$=1.5) and free space:
    \begin{equation}
    M_{Rx_4}=\left(%
\begin{array}{cc}
  1 & 0 \\
  \frac{n_1-n_0}{Rx_4} & 1 \\
\end{array}%
\right),\quad M_{Ry_4}=\left(\begin{array}{cc}
  1 & 0 \\
  \frac{n_1-n_0}{Ry_4} & 1 \\
\end{array}%
\right), \label{L2ref2}
    \end{equation}

and free space propagation over $z$:
    \begin{equation}
    M_{z}=\left(%
\begin{array}{cc}
  1 & z \\
  0 & 1 \\
\end{array}%
\right), \label{z}
    \end{equation}

the whole ABCD system is described.

\section{Appendix B : Amplitude distributions $R\left(x,y\right)$ and $O\left(x,y\right)$}\label{appB}

In Appendix \ref{appA} each part of our optical system has been
represented with transfer matrices. Using this formalism allows us,
under paraxial conditions, to deal with propagation of a Gaussian
point source through the pipe.

Intensity distribution of the diffraction pattern in the CCD sensor
plane is determined by considering two matrix systems: $M_1^{x,y}$
and $M_2^{x,y}$.

$M_1^{x,y}$ is composed of three steps : propagation in free space
over $z_p$, propagation through the first thick lens, propagation in
a medium of refractive index $n_2$ over $\delta$. It is
characterized by two transfer matrices:
\begin{equation}
M_1^{x,y}=M_\delta \times M_{L_1}^{x,y} \times M_{z_p}=\left(%
\begin{array}{cc}
  A^{x,y}_1 & B^{x,y}_1 \\
  C^{x,y}_1 & D^{x,y}_1 \\
\end{array}%
\right). \label{ABCD1}
\end{equation}

$M_2^{x,y}$ is also composed of three steps : propagation in a
medium of refractive index $n_2$ over $z_i$, propagation through the
second thick lens, propagation in free space over $\delta$. Transfer
matrices for this system are:
\begin{equation}
M_2^{x,y}=M_z \times M_{L_2}^{x,y} \times M_{z_i}=\left(%
\begin{array}{cc}
  A^{x,y}_2 & B^{x,y}_2 \\
  C^{x,y}_2 & D^{x,y}_2 \\
\end{array}%
\right). \label{ABCD2}
\end{equation}

Thanks to Eqs. (\ref{L1ref1}), (\ref{L1prop}), (\ref{L1ref2}) one
can build the transfer matrices of the first thick lens:
\begin{equation}
M_{L_1}^x=M_{Rx_2} \times M_e \times M_{Rx_1},\quad
M_{L_1}^y=M_{Ry_2} \times M_e \times M_{Ry_1}. \label{L1}
\end{equation}
By using the same method for the second lens, we obtain:
\begin{equation}
M_{L_2}^x=M_{Rx_4} \times M_e \times M_{Rx_3},\quad
M_{L_2}^y=M_{Ry_4} \times M_e \times M_{Ry_3}. \label{L2}
\end{equation}

    \subsection{Propagation through $M_1^{x,y}$}

After analytical developments of Eq. (\ref{genefresnel}), the
complex amplitude distribution in the object plane is:
\begin{equation}
G_{1}\left(\xi,\eta\right)=\frac{\exp\left(i\frac{2\pi}{\lambda}E_1\right)}{i\lambda
\sqrt{B^x_1B^y_1}}K_1^{x}K_1^{y}\exp\left[-\left(\frac{\xi^2}{\omega_{1_{x}}^2}+\frac{\eta^2}{\omega_{1_{y}}^2}\right)\right]\exp\left[-i\frac{\pi}{\lambda}\left(\frac{\xi^2}{R_{1_{x}}}+\frac{\eta^2}{R_{1_{y}}}\right)\right],
\label{G1}
\end{equation}
with $K_1^{x,y}$ given by:
\begin{equation}
     K_1^{x,y}=\left(\frac{\pi \omega^2}{1-iA^{x,y}_1\frac{\pi \omega^2}{\lambda B^{x,y}_1}}\right)^{1/2},
\end{equation}
$\omega_{1_{x,y}}$ and $R_{1_{x,y}}$ are respectively the beam radii
and the wavefront curvature in the particle plane. Their
mathematical expressions are:
\begin{equation}
    \omega_{1_{x,y}}=\left(\frac{\lambda B^{x,y}_1}{\pi \omega}\right)\left[1+\left(A^{x,y}_1\frac{\pi \omega^2}{\lambda B^{x,y}_1}\right)^2\right]^{1/2},\quad
    R_{1_{x,y}}=-\frac{B^{x,y}_1}{D^{x,y}_1-\frac{A^{x,y}_1\left(\frac{\pi \omega^2}{\lambda B^{x,y}_1}\right)^2}{1+\left(A^{x,y}_1\frac{\pi \omega^2}{\lambda B^{x,y}_1}\right)^2}}.
\end{equation}

    \subsection{Propagation through $M_2^{x,y}$}

Propagation to the CCD sensor plane is calculated thanks to the
generalized Huygens-Fresnel integral (see Eq. (\ref{HF_G2})).

        \subsubsection{Amplitude distribution $R\left(x,y\right)$}

$R$ is associated with the reference wave. After analytical
developments of Eq. (\ref{HF_I1}), the amplitude distribution $R$
is:
\begin{multline}
R\left(x,y\right)=\frac{\exp\left(i\frac{2\pi}{\lambda}E_1\right)}{i\lambda\sqrt{B^x_1B^y_1}}K_1^x K_1^y K_2^x K_2^y\\
\times\exp\left[-\frac{\pi}{\lambda}\left(\frac{N_x}{B^x_2}x^2+\frac{N_y}{B^y_2}y^2\right)\right]\exp\left[i\frac{\pi}{\lambda}\left(\frac{M_x}{B^x_2}x^2+\frac{M_y}{B^y_2}y^2\right)\right],
\end{multline}
where,
\begin{equation}
M_{x,y}=D^{x,y}_2+\frac{\left(\frac{\pi \omega_{1_{x,y}}^2}{\lambda
B^{x,y}_2}\right)^2\left(\frac{B^{x,y}_2}{R_{1_{x,y}}}-A^{x,y}_2\right)}{1+\left(\frac{\pi
\omega_{1_{x,y}}^2}{\lambda
B^{x,y}_2}\right)^2\left(\frac{B^{x,y}_2}{R_{1_{x,y}}}-A^{x,y}_2\right)^2},
\quad N_{x,y}=\frac{\frac{\pi \omega_{1_{x,y}}^2}{\lambda
B^{x,y}_2}}{1+\left(\frac{\pi \omega_{1_{x,y}}^2}{\lambda
B^{x,y}_2}\right)^2\left(\frac{B^{x,y}_2}{R_{1_{x,y}}}-A^{x,y}_2\right)^2},
\end{equation}
and
\begin{equation}
K_2^{x,y}=\left[\frac{\pi \omega_{1_{x,y}}^2}{1+i\frac{\pi
\omega_{1_{x,y}}^2}{\lambda
B^{x,y}_2}\left(\frac{B^{x,y}_2}{R_{1_{x,y}}}-A^{x,y}_2\right)}\right]^{1/2}.
\end{equation}

        \subsubsection{Amplitude distribution $O\left(x,y\right)$}

$O$ is the amplitude of the diffracted wave. We define
$\omega_{{1_{x,y}}_{eq}}$ and $R_{{1_{x,y}}_{eq}}$
\begin{equation}
\frac{1}{\omega_{{1_{x}}_{eq}}^2}=\frac{1}{\omega_{1_{x}}^2}+\frac{\Re\{B_k\}}{b^2},\quad\frac{1}{\omega_{{1_{y}}_{eq}}^2}=\frac{1}{\omega_{1_{y}}^2}+R_{ell}^2\frac{\Re\{B_k\}}{b^2},
\end{equation}
and
\begin{equation}
\frac{1}{R_{{1_{x}}_{eq}}}=\frac{1}{R_{1_{x}}}+\frac{\Im\{B_k\}\lambda}{\pi
b^2},\quad\frac{1}{R_{1{_{y}}_{eq}}}=\frac{1}{R_{1_{y}}}+R_{ell}^2\frac{\Im\{B_k\}\lambda}{\pi
b^2},
\end{equation}
to simplify notations. It should be noted that $\Re$ and $\Im$ stand
for real and imaginary part respectively. Thus $O$ becomes:
\begin{multline}
O\left(x,y\right)=\frac{\exp\left(i\frac{2\pi}{\lambda}E_1\right)}{i\lambda\sqrt{B^x_1B^y_1}}K_1^x K_1^y\exp\left[i\frac{\pi}{\lambda}\left(\frac{D^x_2}{B^x_2}x^2+\frac{D^y_2}{B^y_2}y^2\right)\right]\sum_{k=1}^{N}A_kK_2^{{x}_{eq}}K_2^{{y}_{eq}}\\
\times\exp\left[-\frac{\pi}{\lambda}\left(\frac{N_{x_{eq}}}{B^x_2}x^2+\frac{N_{y_{eq}}}{B^y_2}y^2\right)\right]\exp\left[i\frac{\pi}{\lambda}\left(\frac{M_{x_{eq}}}{B^x_2}x^2+\frac{M_{y_{eq}}}{B^y_2}y^2\right)\right],
\end{multline}
with
\begin{multline}
M_{{x,y}_{eq}}=\frac{\left(\frac{\pi
\omega_{{1_{x,y}}_{eq}}^2}{\lambda
B^{x,y}_2}\right)^2\left(\frac{B^{x,y}_2}{R_{{1_{x,y}}_{eq}}}-A^{x,y}_2\right)}{1+\left(\frac{\pi
\omega_{{1_{x,y}}_{eq}}^2}{\lambda
B^{x,y}_2}\right)^2\left(\frac{B^{x,y}_2}{R_{{1_{x,y}}_{eq}}}-A^{x,y}_2\right)^2},
\\
N_{{x,y}_{eq}}=\frac{\frac{\pi \omega_{{1_{x,y}}_{eq}}^2}{\lambda
B^{x,y}_2}}{1+\left(\frac{\pi \omega_{{1_{x,y}}_{eq}}^2}{\lambda
B^{x,y}_2}\right)^2\left(\frac{B^{x,y}_2}{R_{{1_{x,y}}_{eq}}}-A^{x,y}_2\right)^2},
\end{multline}
and
\begin{equation}
K_2^{{{x,y}_{eq}}}=\left[\frac{\pi
\omega_{1_{{x,y}_{eq}}}^2}{1+i\frac{\pi
\omega_{1_{{x,y}_{eq}}}^2}{\lambda
B^{x,y}_2}\left(\frac{B^{x,y}_2}{R_{1_{{x,y}_{eq}}}}-A^{x,y}_2\right)}\right]^{1/2}.
\end{equation}


\newpage
\listoffigures
\newpage

\end{document}